\documentclass[11pt]{article}
\usepackage{amssymb}
\usepackage{amsmath}
\usepackage{setspace}
\usepackage[left=2.5cm,right=2.5cm,top=3cm]{geometry} 
\newcommand{\cref}[1]{eqn.~(\ref{#1})}
\newcommand{\fn}{\footnote}

\begin{document}
\title{The Second Law in 4D Einstein-Gauss-Bonnet Gravity}
\author{Saugata Chatterjee${}^1$\fn{schatte8@asu.edu} and Maulik Parikh${}^{1,2}$\fn{maulik.parikh@asu.edu} \vspace{0.2in}\\
${}^1$Department of Physics, Arizona State University, \\
Tempe, Arizona 85287, USA \vspace{0.1in} \\
${}^2$Beyond Center for Fundamental Concepts in Science, \\
Arizona State University, Tempe, Arizona 85287, USA\\
}

\date{}

\maketitle

\begin{abstract}
\noindent
The topological contribution to black hole entropy of a Gauss-Bonnet term in four dimensions opens up the possibility of a violation of the second law of thermodynamics in black hole mergers. We show, however, that the second law is not violated in the regime where Einstein-Gauss-Bonnet holds as an effective theory and black holes can be treated thermodynamically. For mergers of AdS black holes, the second law  appears to be violated even in Einstein gravity; we argue, however, that the second law holds when gravitational potential energy is taken into account.
\end{abstract}

\vspace{0.2in}

\noindent
Bekenstein-Hawking entropy \cite{Bekenstein:1973ur}, which is proportional to the surface area of a black hole, always increases in time for classical processes \cite{Hawking:1971vc,Bardeen:1973gs}. This is true even when the black hole is subject to large changes, as it is during black hole mergers \cite{Siino:2009vw}. However, the Bekenstein-Hawking entropy is the correct entropy only if the gravitational sector of the underlying theory is described by the Einstein-Hilbert action; when the action contains higher-order Riemann curvature terms, a different expression for entropy is necessary. For example, Wald entropy \cite{Wald:1993nt} is constructed in order to explicitly satisfy the first law of thermodynamics for black holes in higher-curvature gravity. It remains an open question whether the entropy formulas for event horizons in these more general gravitational theories also obey the second law of thermodynamics. Indeed, it has been argued in \cite{Liko:2007vi} and \cite{Sarkar:2010xp} (see also \cite{Jacobson:1993xs}) that the presence of a Gauss-Bonnet term in the four-dimensional gravitational action should --- on general grounds that are reviewed in this paper --- lead to second law violations during black hole mergers. In this paper, we examine this claim carefully and argue that no violations of the second law can occur in the regime where both Einstein-Gauss-Bonnet holds as an effective theory and black hole thermodynamics is valid.\footnote{Our approach differs from Hawking's proof of the area theorem in three ways. First, we are including a Gauss-Bonnet term. Second, for black hole mergers, we are dealing with the micro-canonical ensemble (fixed total energy) whereas the area theorem applies to the canonical ensemble of fixed temperature. Third, a black hole merger is a topology-changing process rather than a small perturbation. Thus, our demonstration of the validity of the second law is an addition to the pre-existing proofs and does not automatically follow from them.}

Consider the Einstein-Hilbert action with a Gauss-Bonnet term (disregarding surface terms \cite{Myers:1987yn}):
\begin{equation}
I = \frac{1}{16 \pi G}\int d^4 x \sqrt{-g}\left( -2 \Lambda + R + \alpha  (R^2 - 4 R_{ab} R^{ab} + R_{abcd} R^{abcd}) \right) \; . \label{eq:action}
\end{equation}
Here $\alpha$ is a constant with dimensions of $(\mbox{length})^2$. This combination of curvature-squared terms is non-dynamical in four dimensions. One quick way to see this is to Wick-rotate the action. Then the  Euclideanized Gauss-Bonnet term integrated over a compact four-surface is 
simply proportional to the Euler character of that surface: 
\begin{equation}
\chi_4 = \frac{1}{32 \pi^2} \int  d^4 x \sqrt{(g)_E} \left(  R^2 - 4 R_{ab} R^{ab} + R_{abcd} R^{abcd} \right)_{\! E} \; . \label{eq:chi-4}
\end{equation}
Because it is topological, the Gauss-Bonnet term in four dimensions does not affect Einstein's equations, and therefore, when Wick-rotated back, this action has the same class of black hole solutions as pure Einstein gravity. Nevertheless, even though it does not contribute to the equations of motion, the extra Gauss-Bonnet contribution does have ramifications for semi-classical gravity (see, e.g. \cite{Parikh:2009js}), because it alters the definition of entropy. In Lovelock gravity, the entropy is not given by the area but by the Jacobson-Myers entropy formula \cite{Jacobson:1993xs}; modification of the entropy formula is necessary for the validity of the first law. Jacobson-Myers entropy is suitable for dynamical black hole horizons similar to the boost-invariant form of the Wald entropy \cite{Iyer:1994ys}. Since we are attempting to study black hole mergers, this entropy is preferable in form to the Wald entropy \cite{Wald:1993nt} which assumes the stationarity of horizons. The Jacobson-Myers entropy associated with arbitrary black holes of 4D Einstein-Gauss-Bonnet theory \cite{Jacobson:1993xs} is
\begin{equation}
S = \frac{1}{4 G} \int dA  \left( 1 + \alpha R_{(2)} \right)  \label{eq:ent} \; .
\end{equation}
Here $R_{(2)}$ is the Ricci scalar of a two--dimensional  (2D) spacelike section of the event horizon.
Integrating \cref{eq:ent} is straightforward because the last part of the integral is simply the Euler character of the surface of the black hole:
\begin{equation}
\chi = \frac{1}{4 \pi} \int  dA \, R_{(2)}  \label{eq:chi-2} \; .
\end{equation}
Since this integral is over a spacelike hypersurface, it directly gives the Euler number without any Euclideanization; also, $\chi$ is the 2D Euler character, which should not be confused with $\chi_4$ in \cref{eq:chi-4}.) The entropy is now the sum of two terms -- the usual Bekenstein-Hawking term plus an additional term proportional to the Euler number:
\begin{equation}
S = \frac{A}{4 G} + \pi \frac{\alpha }{G} \chi \label{eq:entropy} \; .
\end{equation}
At first glance, this formula opens the door to second law violations when $\alpha > 0$. To see this, consider the merger of two black holes with the spherical topology ($\chi = 2$). For each black hole the Gauss-Bonnet term contributes $\frac{2 \pi \alpha}{G}$ to the entropy, but after the merger only one hole exists. Thus, it might be that the increase in area entropy could be outweighed by the decrease in the topological contribution to the entropy.
However, we need to keep in mind two regimes of validity. Since gravity is not a renormalizable theory, the effective action consists of an infinite number of terms of ever-higher order in powers of the Riemann tensor. For example, the Einstein-Gauss-Bonnet action appears as only the leading terms in the low-energy effective action of heterotic string theory \cite{Zwiebach:1985uq,Gross:1986iv}. In order to be able to neglect higher-order terms, a necessary condition is that $O(|\alpha| R) < 1$, where $R^2$ denotes some quadratic curvature scalar.  Just on dimensional grounds, we see that the largest value the curvature scalar can have is of the order $R \sim 1/\ell_P^2 \sim M_P^2 \sim 1/G$, which implies that $|\alpha|/G < 1$ \fn{$\ell_P$ is the Planck length and  $M_P$ is the Planck mass. In natural units, $\ell_P = 1/M_P$ and $\frac{1}{16 \pi G} = \frac{M_P^2}{2} $}. Moreover, in order for a semi-classical thermodynamic description to be valid, at least one of the merging black holes must have a large entropy (i.e., it must be macroscopic: $M \gg M_P$. Thus, any semi-classical treatment of black holes in Einstein-Gauss-Bonnet gravity assumes the validity of two conditions:
\begin{equation}
\frac{|\alpha|}{G} < 1	\label{EFT}
\end{equation}
\begin{equation}
\frac{M}{M_P} \gg 1	\label{ClassBH}
\end{equation}
With these conditions in mind, let us attempt to force a violation of the second law in a merger. This can be attempted for black holes in asymptotically Minkowski, de Sitter, and anti-de Sitter space, depending on the value of any cosmological constant. We treat these three cases in turn.

%%%%%%%%%%%%%%%%%%%%%%%%%%%%%%%%%%%%%%%%%%%%%%%%%%%%%%%%%%%%%%%%%%%%%
\subsection*{Asymptotically flat black hole spacetimes}\label{sec:asymp-flat}

Since we are attempting to engineer a violation of the second law, let us first identify a scenario in which the increase in area is minimal, since any area increase contributes positively to the change in entropy. For a given mass, extremal black holes have the smallest area, $A = 4 \pi G^2 M^2$ (compared with, say, $16 \pi G^2 M^2$ for a Schwarzschild black hole). Let us therefore consider the merger of two extremal black holes, neglecting the loss due to gravitational waves emitted. Extremal black hole solutions with the same charge, known as Majumdar-Papapetrou black holes \cite{Hartle:1972ya}, have no mutual forces, and hence solutions of single black holes can be superimposed to give exact multi centered solutions. In this case, the entropy of the collection of black holes is just the sum of the entropies of each individual black hole.
The Bekenstein-Hawking entropy of a single extremal black hole in pure Einstein gravity is $S = \pi G M^2$. If two extremal black holes of masses $M$ and $M'$ merge, the net change in the area entropy is therefore $\Delta S = 2 \pi G M M'$. 
For a macroscopic $M$, the smallest possible increase in area entropy occurs when the second black hole has a mass of $M_P$.
Then the change in area entropy is $\Delta S = 2 \pi G M_P M $. Let us try to offset this by including the Gauss-Bonnet contribution. 
The initial entropy of the system is
\begin{eqnarray}
S_i = \pi G ( M^2 + M_P^2) + \frac{4 \pi \alpha}{G} \; .
\end{eqnarray}
The entropy after merger is
\begin{eqnarray}
S_f = \pi G ( M + M_P)^2 + \frac{2 \pi \alpha}{G} \; .
\end{eqnarray}
The change in entropy is then
\begin{equation}
\Delta S = 2 \pi G M_P M - 2 \pi \frac{\alpha}{G} \; . \label{DeltaS}
\end{equation}
For a violation of the second law to occur, we require
\begin{equation}
\frac{\alpha}{G} > G M_P M = \left( \frac{M}{M_P} \right) \; .   \label{eq:a-bound-1}
\end{equation}
However, this requirement contradicts eqns. (\ref{EFT}) and (\ref{ClassBH}). Thus, in order for a second law violation to take place in Einstein-Gauss-Bonnet gravity, either we must have $M/M_P < 1$, in which case the ``black hole" has no description in terms of classical geometry, or we must have $\alpha/G \gg 1$, invalidating Einstein-Gauss-Bonnet as an effective theory of gravity.
 
Although the coefficient $\alpha$ is positive in string theory, let us briefly consider the consequences of negative $\alpha$. When $\alpha < 0$, the merger process actually causes entropy to increase, by eqn. (\ref{DeltaS}). However, now we have to check that the entropy of even one hole is positive. For the holes to have positive entropy,
\begin{equation}
\frac{|\alpha|}{G} < \frac{1}{2} \left( \frac{M}{M_P}\right)^{\! 2}  \; .  \label{eq:a-bound-2}
\end{equation}
In our regime of validity, eqns. (\ref{EFT}) and (\ref{ClassBH}) are obeyed, this bound is automatically satisfied. Thus, the negative $\alpha$ case presents no problems insofar as black hole thermodynamics is concerned.

For more general (e.g. Kerr) black holes, there are no exact, stable two-black-hole solutions. However, the area entropy of such black holes is greater than that of the Majumdar-Papapetrou black holes we considered. Therefore, one expects that in a merger of Kerr black holes, the entropy should increase even more.
%%%%%%%%%%%%%%%%%%%%%%%%%%%%%%%%%%%%%%%%%%%%%%%%%%%%%%%%%%%%%%%%%%%%%

\subsection*{Asymptotically de Sitter black hole spacetimes}\label{sec:asymp-AdS}

Consider next black hole mergers in asymptotically de Sitter space. If the two black holes are both much smaller than the de Sitter scale, the de Sitter curvature scale becomes irrelevant and our results for asymptotically flat black holes apply. Alternatively, if both black holes are large, we cannot merge them. This is because in de Sitter space, there is a maximum mass black hole, the Nariai solution, with $GM_{\rm max} = L/\sqrt{27}$, where $L$ is the de Sitter length. Hence we cannot merge two black holes whose combined mass exceeds the Nariai mass. Moreover, even if the total mass is less than $M_{\rm max}$, large black holes cannot be separated in a manner where we can reliably add their entropies. The only case left consists of  one black hole has large mass and another with infinitesimal mass. Consider then a black hole with mass $M_{\rm max} - M_P$ and a black hole of mass $M_P$, so that the combined mass is the Nariai mass (for simplicity). The horizon of the large mass black hole has radius $r_1 = \frac{L}{\sqrt{3}} \left( 1 - \epsilon- \frac{1}{6} \epsilon^2 \right)$, where $\epsilon \equiv \sqrt{2M_P/3M} \ll 1$ \cite{Ginsparg:1982rs}, while the small mass black hole has radius $r_2 = 2 G M_P$. The final configuration has only a Nariai black hole. Since the total mass is fixed, the cosmological horizon does not change during the merger and we can neglect the entropy contribution of the cosmological horizon. Considering only the black hole horizons, the Nariai black hole  has entropy $\frac{\pi L^2}{3G} + \frac{2 \pi \alpha}{G}$. On the other hand, the entropy of the large black hole and small black hole system is $\frac{\pi}{G} (r_1^2) + 4 \pi G M_P^2 + \frac{4 \pi \alpha}{G} \simeq  \frac{\pi L^2}{3 G}\left( 1  -2  \epsilon \right) + 4 \pi G M_P^2 + \frac{4 \pi \alpha}{G}$. Note that the entropic contribution of the microscopic black hole and the Gauss-Bonnet terms are both sub-leading in $\epsilon$. Hence, to leading order in $\epsilon$, the change in entropy is
\begin{eqnarray}
\Delta S &\simeq & \pi G \sqrt{216 M_P M^3} \; ,
\end{eqnarray}
which is obviously positive. This result is consistent with earlier results \cite{Liko:2012wq}.

%%%%%%%%%%%%%%%%%%%%%%%%%%%%%%%%%%%%%%%%%%%%%%%%%%%%%%%%%%%%%%%%%%%%%
\subsection*{Asymptotically anti-de Sitter black hole spacetimes}\label{sec:asymp-AdS}

One of the features of asymptotically AdS spaces is that they allow black hole solutions with non-compact horizons. Even though the ``uniqueness theorem" \cite{Israel:1967wq} dictates that the horizon topology of asymptotically flat 4D black holes must be spherical, no such restrictions apply for black holes in asymptotically AdS spaces; flat and hyperbolic horizon topologies are also allowed \cite{Mann:1996gj, Cai:1996eg, Brill:1997mf, Vanzo:1997gw, Smith:1997wx, Aminneborg:1996iz}. The generalization of the Schwarzschild solution is
\begin{equation}
ds^2 = - f dt^2 + f^{-1} dr^2 + r^2 d\Sigma_{2}^2 \; , \label{eq:hp-bh}
\end{equation}
where
\begin{equation}
f = k - \frac{2GM}{r} + \frac{r^2}{L^2} \; .
\end{equation}
Here $d\Sigma_{2}^2$ is the line element of the spacelike section of the horizon with constant curvature, $k = +1,0,-1$, corresponding to a positive-, zero, or negative-curvature horizon respectively. The mass of the black hole is $M$, which is obtained by the Abbott-Deser-Tekin formalism for asymptotically AdS spaces \cite{Abbott:1981ff,Deser:2002jk}. The $k = -1,0$ black holes have infinite area but can be made compact by suitable identifications (e.g. \cite{Parikh:2005qs}). In the hyperbolic case ($k = -1$), identification allows for horizons with different spatial topologies; spacelike sections with genus $g > 1$ are isomorphic to the quotient space of 2D hyperbolic space under discrete isometries.
It is certainly interesting that the boundary contribution of the variation of the bulk metric for asymptotically AdS spaces can generate the bulk Gauss-Bonnet term. However, this only fixes the Gauss-Bonnet coupling if there is no Gibbons-Hawking term and so it is somewhat tangential to the main idea here.

%%%%%%%%%%%%%%%%%%%%%%%%%%%%%%%%%%%%%%%%%%%%%%%%%%%%%%%%%%%%%%%%
\subsubsection*{Compact black holes in AdS}

First we shall consider the merger of black holes with compact horizons, one of which is the solution to \cref{eq:hp-bh} with $f(r) = 1 - 2GM/r + r^2/L^2$. However, black holes in AdS which have small masses are thermodynamically unstable by the Hawking-Page transition \cite{Hawking:1982dh}. Therefore, we need to consider only large mass black holes. The horizon is at 
\begin{equation}
  r_h = L^2/3Q + Q ; Q =  \left( G M L^2  + \sqrt{ G^2 M^2 L^4 + L^6/27}\right)^{1/3} \; .
\end{equation}
In the large-mass limit, $GM \gg L$. The horizon radius becomes $r_h \approx (2GML^2)^{1/3} \left( 1 + \frac{1}{3} \left( \frac{L}{2G M}\right)^{2/3} \right)$. Including the Gauss-Bonnet term in the entropy and setting $\chi=2$ for the compact horizon, we find
\begin{equation}
S =  \frac{A}{4 G} + \pi \frac{\alpha }{G} \simeq \frac{\pi}{G} \left( (2GML^2)^{2/3} + \frac{2}{3} L^2  + 2\alpha  \right) \; . \label{eq:entropy-ads-bh-1}
\end{equation}
The entropy can potentially be rendered negative by the last Gauss-Bonnet term if the coupling constant, $\alpha$, is negative. However, for effective field theory to be valid, the Gauss-Bonnet term in the action, eqn. (\ref{eq:action}), must be much smaller than the preceding terms. This means in particular that $\alpha R^2 \ll \Lambda \Rightarrow \alpha/L^4 \ll 1/L^2$.  Therefore we have 
\begin{equation}
\frac{|\alpha|}{L^2} \ll 1 \; .
\end{equation}
In view of this constraint, it is easy to see that the entropy is always positive.  

Having established the positivity of the entropy, let us consider the merger, for simplicity, of two equal-mass AdS black holes. The change in the entropy due to this merger process would be
\begin{eqnarray}
\Delta S \simeq \frac{\pi}{G} \left((2^{2/3} -2) (2GML^2)^{2/3} - \frac{2}{3} ~L^2 - 2 \alpha \right) \; .
\end{eqnarray}
In the limit that we are working, $G M \gg L$, the first two terms always add up to a negative value, and the last Gauss-Bonnet term just makes things worse when $\alpha>0$. Thus it appears that, even in Einstein gravity, the merger of AdS black holes appears to violate the second law.

The resolution is as follows. Our approach has been to compare the final entropy of one black hole, given by an exact solution, to the initial entropies of two black holes. Except in the Majumdar-Papapetrou case, the two-black-hole geometries are not exact, stable solutions of general relativity. To get the entropy, we have added the entropies of single-black-hole solutions. That is acceptable if the presence of each black hole is only a small perturbation on the geometry near the other black hole. In asymptotically flat space, we could consider two widely separated black holes. When $2 GM/r \ll 1$, the metric near each black hole would be independent of the existence of the other black hole. In order to be able to add entropies, our initial configuration has to have a minimum separation of the holes. Now, in asymptotically AdS space, the minimum separation depends on the AdS scale too: $2GM/r \ll  r^2/L^2$. This means that, in the background and coordinates of an AdS black hole of mass $M_1=M$, the other black hole of mass $M_2$ must be located at least at $r_0 \gg (2 G M L^2)^{1/3} \simeq \eta (2 G M L^2)^{1/3}$, where $\eta$ is a large number. However, in AdS, this requires climbing up a potential energy barrier. One can estimate how much energy is required to separate black holes to our minimum separation by considering a point particle in the geometry of the other. 
Let the energy of the black hole be $M_2 = -p_0$, where $p^\mu$ is the four-momentum of the black hole (we can treat it as a particle by ignoring its back reaction on the background). For a particle of mass $M$ at rest at a radius, $r_0$, in an AdS space, $p_0 = - M\sqrt{1 + r_0^2/L^2}$, which follows from $-M^2 = p_0^2 g^{00}$. Hence, when the second black hole is brought from $r=r_0$ to $r=0$, the total energy of the combined black hole system is not just the sum of the two masses, but must also include this potential energy:

\begin{eqnarray}
M_{\rm tot} = M_1 + M_2= M +  M\sqrt{1 + r_0^2/L^2} \simeq M \left( 1 + \frac{r_0}{L} \right) \simeq \eta \left( \frac{2G}{L}\right)^{\! 1/3} M^{4/3} \; .
\end{eqnarray}
When we compare the entropy of the resultant black hole, we find that it exceeds the sum of the initial entropies:
\begin{eqnarray}
  \Delta S =  S(M_{\rm tot}) - 2 S(M) \simeq \frac{\pi}{G} \left( (2GL^2)^{2/3}  \left(  \left( \frac{2 \eta ^3 G}{L} \right)^{\! 2/9} M^{8/9} - 2 M^{2/3} \right)  - \frac{2}{3} L^2 \right) \; . \label{eq:delta-S-1}
\end{eqnarray} 
The term in brackets is positive because $M > \frac{2^{7/2}L}{\eta^3 G}$ always, since $GM \gg L$ and $\eta \gg 1$. Also, it is of the order $M^{2/3}$, implying that  $ \Delta S \sim (G ML^2)^{2/3} - L^2 $, and since $G M$ is large compared to $L$, $\Delta S$ is positive.

Now we can add the Gauss-Bonnet  contribution to the entropy. Taking minimum mass black holes so that the $M_{\rm tot}^{2/3} - 2 M^{2/3} \sim L^2$,
\begin{eqnarray}
 \Delta S \simeq  \frac{\pi}{G} \left( L^2  - 2 \alpha  \right) \; .
\end{eqnarray}
However, as argued earlier, the coupling constant, $\alpha$, satisfies the constraint $|\alpha|/L^2 < 1$. Therefore, $\Delta S > 0$ even with the Gauss-Bonnet contribution to the entropy.
%%%%%%%%%%%%%%%%%%%%%%%%%%%%%%%%%%%%%%%%%%%%%%%%%%%%%%%%%%%%%%%%
\subsubsection*{Non-compact black holes in AdS}

AdS also admits black holes with non compact horizons.  Consider a hyperbolic black hole with $f = -1 - 2GM/r + r^2/L^2$. The horizon is at
\begin{eqnarray}
r_h = L^2/3Q + Q ;Q = \left( GM L^2 + \sqrt{G^2 M^2 L^4  - L^6/27}\right)^{1/3} \; .
\end{eqnarray}
One important thing to note from this expression is that the mass of a hyperbolic black hole in AdS has a minimum value. It has to satisfy $G M \geq L/\sqrt{27}$. To obtain a finite entropy, we have to compactify the horizon by making discrete identifications.
In the large-mass limit, the total entropy of the compactified horizon of genus $g > 1$ is
\begin{eqnarray}
S &\simeq& \frac{A_0}{4G} \left( (2GML^2)^{2/3} +  \frac{2}{3} L^2 \right)  + \pi \frac{\alpha}{G} \chi \nonumber \\
    &\simeq& \frac{\pi |\chi|}{G} \left(  \left(  \frac{GML^2}{2} \right)^{\! 2/3} + \frac{1}{3} L^2  - \alpha \right) \; , \label{eq:hb-entropy}
\end{eqnarray}
where $A_0$ is the dimensionless area of the compact, orientable horizon of genus $(> 1) = -2\pi\chi$ (by the Gauss-Bonnet theorem) which is positive since $\chi < 0$ for $g > 1$. Here we do not need to worry about the positivity of entropy as in \cref{eq:entropy-ads-bh-1} because the first two terms are order $\sim L^2$ and we have already seen that the coupling constant, $\alpha$, satisfies the constraint $|\alpha|/L^2 \ll 1$. 

After merging two such hyperbolic black holes of equal mass and ignoring the Gauss-Bonnet contribution for the moment, the change in entropy is $\Delta S \simeq (\pi|\chi|/G)( 2^{-1/3} (2^{2/3} -2)~ (GML^2)^{2/3} - \frac{1}{3} ~L^2) $, which is negative. However, we have again neglected the effect of the potential energy gained by the black holes while coming from a large distance in AdS. An analysis similar to that for compact AdS black holes leads to a change in entropy of the form
\begin{eqnarray}
  \Delta S  \simeq \frac{\pi |\chi|}{G}  \left( \left( \frac{G^2 L^4}{2}\right)^{\! 1/3} \left(   \left( \frac{2G\eta^3}{L} \right)^{\! 2/9} M^{8/9} - 2 M^{2/3} \right)   - \frac{1}{3} ~L^2 + \alpha\right) \; .
\end{eqnarray}  
Similar to the previous case of compact black holes, a black hole mass $M > M_P$ is enough for $\Delta S$ to be positive. The Gauss-Bonnet term again cause no trouble due to its smallness compared to the AdS scale.

For planar black holes, with $f = - 2GM/r + r^2/L^2$, the analysis is similar. A finite entropy can be obtained by making a toroidal identification on the plane. Again, a proper accounting of the potential energy ensures that the entropy increases in a merger. The Gauss-Bonnet term has no effect here since the Euler character of a torus is zero.

%%%%%%%%%%%%%%%%%%%%%%%%%%%%%%%%%%%%%%%%%%%%%%%%%%%%%%%%%%%%%%%%%%%%%
\subsection*{Conclusion}

We have investigated the validity of the second law of thermodynamics for black hole mergers in 4D Einstein-Gauss-Bonnet effective theory. Contrary to previous claims in the literature, we see that the second law remains valid within the regime of validity of approximations, even though the presence of a topological term threatens to decrease the total entropy. Our calculations here are not at the level of a proof, and we did not consider the most general 4D black hole. But we have shown that the reasoning suggesting that the second law is violated does not apply, and we see no reason to suspect second law violation for more general black holes. Nevertheless, it would be worth examining mergers of other types of black holes.

Of course, the second law is one of the most robust laws in physics: in any theory with consistent underlying statistical mechanics, coarse-grained entropy is expected to increase when two macroscopic systems merge. Had we found a violation of the second law for black holes in Einstein-Gauss-Bonnet gravity, it would have called into question not so much the second law as a principle of nature, but the semi-classical consistency of Einstein-Gauss-Bonnet theory. Our results show that the second law is indeed obeyed; we regard this as evidence that the thermodynamics of Einstein-Gauss-Bonnet gravity is consistent with some underlying statistical mechanics.

%%%%%%%%%%%%%%%%%%%%%%%%%%%%%%%%%%%%%%%%%%%%%%%%%%%%%%%%%%%%%%%%%%%%%
\subsection*{Acknowledgement}

We would like to thank David Lowe for useful discussions. M. P. is supported in part by DOE grant DE-FG02-09ER41624.

%%%%%%%%%%%%%%%%%%%%%%%% bibliography %%%%%%%%%%%%%%%%%%%%%%%%%%%%%%

\end{document}